\documentclass[english,prd,superscriptaddress,nofootinbib,preprintnumbers,notitlepage]{revtex4}
\usepackage[T1]{fontenc}
\usepackage[latin1]{inputenc}
\usepackage{bm}
\usepackage{amsmath}
\usepackage{amssymb}
\usepackage{graphicx}
\usepackage{esint}

\makeatletter
\@ifundefined{textcolor}{}
{%
 \definecolor{BLACK}{gray}{0}
 \definecolor{WHITE}{gray}{1}
 \definecolor{RED}{rgb}{1,0,0}
 \definecolor{GREEN}{rgb}{0,1,0}
 \definecolor{BLUE}{rgb}{0,0,1}
 \definecolor{CYAN}{cmyk}{1,0,0,0}
 \definecolor{MAGENTA}{cmyk}{0,1,0,0}
 \definecolor{YELLOW}{cmyk}{0,0,1,0}
 }

\usepackage{amsthm}\usepackage{bm}

\@ifundefined{textcolor}{}{%
 \definecolor{BLACK}{gray}{0}
 \definecolor{WHITE}{gray}{1}
 \definecolor{RED}{rgb}{1,0,0}
 \definecolor{GREEN}{rgb}{0,1,0}
 \definecolor{BLUE}{rgb}{0,0,1}
 \definecolor{CYAN}{cmyk}{1,0,0,0}
 \definecolor{MAGENTA}{cmyk}{0,1,0,0}
 \definecolor{YELLOW}{cmyk}{0,0,1,0}
 }

\def\Mpl{M_{\rm pl}}

\usepackage{babel}

\makeatother

\usepackage{babel}
\begin{document}

\title{Conditions for the cosmological viability of the most general scalar-tensor
theories and their applications to extended Galileon dark energy models}

\author{Antonio De Felice}

\affiliation{TPTP \& NEP, The Institute for Fundamental Study, Naresuan University,
Phitsanulok 65000, Thailand}

\affiliation{Thailand Center of Excellence in Physics, Ministry of Education,
Bangkok 10400, Thailand}

\selectlanguage{english}%

\author{Shinji Tsujikawa}

\affiliation{Department of Physics, Faculty of Science, Tokyo University of Science,
1-3, Kagurazaka, Shinjuku-ku, Tokyo 162-8601, Japan}

\date{\today}
\begin{abstract}
In the Horndeski's most general scalar-tensor theories with second-order
field equations, we derive the conditions for the avoidance of ghosts
and Laplacian instabilities associated with scalar, tensor, and vector
perturbations in the presence of two perfect fluids on the flat Friedmann-Lema\^{i}tre-Robertson-Walker
(FLRW) background. Our general results are useful for the construction
of theoretically consistent models of dark energy. We apply our formulas
to extended Galileon models in which a tracker solution with an equation
of state smaller than $-1$ is present. We clarify the allowed parameter
space in which the ghosts and Laplacian instabilities are absent and
we numerically confirm that such models are indeed cosmologically
viable. 
\end{abstract}
\maketitle

\section{Introduction}

The problem of dark energy is one of the most serious problems faced
by cosmologists and particle physicists. The cosmological constant
is a simplest candidate, but it is plagued by the huge difference
of energy scales between the theoretical and observed values \cite{Weinberg}.
Many alternative models to the cosmological constant have been proposed
to approach the origin of dark energy \cite{review}.

Dark energy models can be broadly classified into two classes: (i)
modified matter models, and (ii) modified gravity models. In the class
(i) the modified matter source is introduced to drive the late-time
cosmic acceleration. The representative model in this class is quintessence
\cite{quin}, in which the potential energy $V(\phi)$ of a slow-rolling
scalar field $\phi$ is the source for dark energy. Another model
is k-essence \cite{kes}, in which a non-linear term with respect
to the field kinetic energy $X=-(\partial\phi)^{2}/2$ leads to the
cosmic acceleration.

The representative models that belong to the class (ii) are those
based on $f(R)$ theories \cite{fR,fRviable}, $f(R,{\cal G})$ theories
\cite{fRG}, Brans-Dicke theories \cite{Brans}, Dvali-Gabadazde-Porrati
(DGP) braneworld \cite{DGP}, and Galileon gravity \cite{Nicolis,DEV}
(see Ref.~\cite{reviewmo} for reviews). The Lagrangian in $f(R)$
theories is an arbitrary function of the Ricci scalar $R$, which
corresponds to the particular class of Brans-Dicke theories (Brans-Dicke
parameter $\omega_{{\rm BD}}=0$ \cite{Ohanlon}). The $f(R,{\cal G})$
theories consist of a function of $R$ and the Gauss-Bonnet term ${\cal G}$,
in which case there are two independent scalar degrees of freedom
in a general background \cite{fRG2f}. In Galileon gravity, the Lagrangian
is constructed such that the field equations are invariant under the
Galilean symmetry $\partial_{\mu}\phi\to\partial_{\mu}\phi+b_{\mu}$
in the limit of Minkowski spacetime \cite{Nicolis}. One of those
terms is given by $X\,\square\phi$, which appears in the DGP model
as a result of the mixing of a brane-bending mode with a transverse
graviton \cite{DGPnon}. In the Galileon model there exists a stable
late-time de Sitter solution where $\dot{\phi}=$\,constant \cite{DePRL,DeTprd}.
Moreover the model can be consistent with local gravity constraints
through the Vainshtein mechanism \cite{Vainshtein}.

Recently Deffayet \textit{et al.} \cite{DGSZ} derived the action
for the most general scalar-tensor theories having second-order field
equations. This action is equivalent to that derived by Horndeski
\cite{Horndeski} in 1974 in the context of Lovelock gravity \cite{KYY}
(see also Ref.~\cite{Char}). In fact the Horndeski's action is sufficiently
general to accommodate most of the dark energy models proposed in
literature. Moreover, the Gauss-Bonnet couplings $F(\phi){\cal G}$
\cite{GBpapers}, the generalized Galileon term $G(\phi,X)\square\phi$
\cite{Koyama,braiding,Kimura,Kimura2}, the derivative couplings $G^{\mu\nu}\partial_{\mu}\phi\partial_{\nu}\phi$
with the Einstein tensor $G^{\mu\nu}$ \cite{Germani} also belong
to the class of the Horndeski's action \cite{KYY}.

For such general theories the full background and perturbation equations
were recently derived in Ref.~\cite{DKT} in the presence of non-relativistic
matter. In the context of inflation several authors computed the power
spectra of scalar and tensor perturbations \cite{KYY} as well as
the non-linear parameter of primordial non-Gaussianities \cite{nongau1,nongau2,nongau3}.

If we modify gravity from General Relativity we need to worry about
the presence of ghosts as well as Laplacian instabilities. In the
presence of two perfect fluids (radiation and non-relativistic matter)
we shall derive conditions for the avoidance of ghosts and Laplacian
instabilities associated with scalar, tensor, and vector perturbations
in the Horndeski's theories. This is of particular importance for
the construction of theoretically consistent modified gravitational
models of dark energy. We apply our results to extended Galileon models
in which a tracker solution is present. This tracker corresponds to
the generalization of that found in Ref.~\cite{DePRL} for the covariant
Galileon. Unlike the covariant Galileon the equation of state of dark
energy $w_{{\rm DE}}$ can take the values close to $-1$ during the
evolution from the matter era to the accelerated epoch. This property
should allow for the tracker to be compatible with observations.

This paper is organized as follows. In Sec.~\ref{backsec} the background
equations of motion are derived on the flat FLRW spacetime in the
presence of two perfect fluids. In Sec.~\ref{avoidancesec} we find
conditions for the absence of ghosts and Laplacian instabilities by
deriving the second-order action for scalar, tensor, and vector perturbations.
In Sec.~\ref{applicationsec} we apply our general formulas to kinetically
driven dark energy models, which cover the covariant Galileon as a
specific case. Not only we identify a theoretically consistent parameter
space but also we numerically integrate the field equations to confirm
the analytic estimation. Sec.~\ref{consec} is devoted to conclusions.

\section{The background equations of motion}

\label{backsec} In order to avoid the Ostrogradski instability \cite{Ost}
it is desirable to keep the equations of motion at second order in
derivatives. The most general 4-dimensional scalar-tensor theories
having second-order field equations are described by the Lagrangian
\cite{DGSZ} 
\begin{equation}
{\cal L}=\sum_{i=2}^{5}{\cal L}_{i}\,,\label{Lagsum}
\end{equation}
 where 
\begin{eqnarray}
{\cal L}_{2} & = & K(\phi,X),\label{eachlag2}\\
{\cal L}_{3} & = & -G_{3}(\phi,X)\Box\phi,\\
{\cal L}_{4} & = & G_{4}(\phi,X)\, R+G_{4,X}\,[(\Box\phi)^{2}-(\nabla_{\mu}\nabla_{\nu}\phi)\,(\nabla^{\mu}\nabla^{\nu}\phi)]\,,\\
{\cal L}_{5} & = & G_{5}(\phi,X)\, G_{\mu\nu}\,(\nabla^{\mu}\nabla^{\nu}\phi)-\frac{1}{6}\, G_{5,X}\,[(\Box\phi)^{3}-3(\Box\phi)\,(\nabla_{\mu}\nabla_{\nu}\phi)\,(\nabla^{\mu}\nabla^{\nu}\phi)+2(\nabla^{\mu}\nabla_{\alpha}\phi)\,(\nabla^{\alpha}\nabla_{\beta}\phi)\,(\nabla^{\beta}\nabla_{\mu}\phi)]\,.\label{eachlag5}
\end{eqnarray}
 Here $K$ and $G_{i}$ ($i=3,4,5$) are functions with respect to
a scalar field $\phi$ and its kinetic energy $X=-\partial^{\mu}\phi\partial_{\mu}\phi/2$,
$R$ is the Ricci scalar, and $G_{\mu\nu}$ is the Einstein tensor.
$G_{i,X}$ and $G_{i,\phi}$ ($i=3,4,5$) correspond to the partial
derivatives of $G_{i}$ with respect to $X$ and $\phi$ respectively,
i.e. $G_{i,X}\equiv\partial G_{i}/\partial X$ and $G_{i,\phi}\equiv\partial G_{i}/\partial\phi$.
The above Lagrangian was first discovered by Horndeski in a different
form \cite{Horndeski}, which is equivalent to the Lagrangian (\ref{Lagsum})
\cite{KYY}.

In addition to the scalar field $\phi$ we take into account two barotropic
perfect fluids, whose energy densities are $\rho_{A}$ and $\rho_{B}$.
Then the total action is given by 
\begin{equation}
S=\int d^{4}x\sqrt{-g}\left({\cal L}+{\cal L}_{A}+{\cal L}_{B}\right)\,,\label{action}
\end{equation}
 where $g$ is the determinant of the metric $g_{\mu\nu}$, ${\cal L}_{A}$
and ${\cal L}_{B}$ are the Lagrangians of two perfect fluids respectively.

Let us consider a flat FLRW background with the metric $ds^{2}=-N^{2}(t)dt^{2}+a^{2}(t)d{\bm{x}}^{2}$,
where $t$ is the cosmic time, $N(t)$ is the lapse function, and
$a(t)$ is the scale factor. Varying the action (\ref{action}) with
respect to the $N(t)$ and $a(t)$ respectively, we obtain 
\begin{eqnarray}
 &  & 2XK_{,X}-K+6X\dot{\phi}HG_{3,X}-2XG_{3,\phi}-6H^{2}G_{4}+24H^{2}X(G_{4,X}+XG_{4,XX})-12HX\dot{\phi}\, G_{4,\phi X}-6H\dot{\phi}\, G_{4,\phi}\nonumber \\
 &  & +2H^{3}X\dot{\phi}\left(5G_{5,X}+2XG_{5,XX}\right)-6H^{2}X\left(3G_{5,\phi}+2XG_{5,\phi X}\right)=-\rho_{A}-\rho_{B}\,,\label{be1}\\
 &  & K-2X(G_{3,\phi}+\ddot{\phi}\, G_{3,X})+2(3H^{2}+2\dot{H})G_{4}-12H^{2}XG_{4,X}-4H\dot{X}G_{4,X}-8\dot{H}XG_{4,X}-8HX\dot{X}G_{4,XX}\nonumber \\
 &  & +2(\ddot{\phi}+2H\dot{\phi})G_{4,\phi}+4XG_{4,\phi\phi}+4X(\ddot{\phi}-2H\dot{\phi})G_{4,\phi X}-2X(2H^{3}\dot{\phi}+2H\dot{H}\dot{\phi}+3H^{2}\ddot{\phi})G_{5,X}-4H^{2}X^{2}\ddot{\phi}\, G_{5,XX}\nonumber \\
 &  & +4HX(\dot{X}-HX)G_{5,\phi X}+2[2(\dot{H}X+H\dot{X})+3H^{2}X]G_{5,\phi}+4HX\dot{\phi}\, G_{5,\phi\phi}=-p_{A}-p_{B}\,,\label{be2}
\end{eqnarray}
 where a dot represents a derivative with respect to $t$, $H\equiv\dot{a}/a$
is the Hubble parameter, and $p_{A},p_{B}$ are the pressures of the
two perfect fluids. Varying the action (\ref{action}) with respect
to $\phi(t)$, it follows that 
\begin{equation}
\frac{1}{a^{3}}\frac{d}{dt}\left(a^{3}J\right)=P_{\phi}\,,\label{fieldeq}
\end{equation}
 where 
\begin{eqnarray}
J & \equiv & \dot{\phi}K_{,X}+6HXG_{3,X}-2\dot{\phi}\, G_{3,\phi}+6H^{2}\dot{\phi}(G_{4,X}+2XG_{4,XX})-12HXG_{4,\phi X}\nonumber \\
 &  & +2H^{3}X(3G_{5,X}+2XG_{5,XX})-6H^{2}\dot{\phi}(G_{5,\phi}+XG_{5,\phi X})\,,\\
P_{\phi} & \equiv & K_{,\phi}-2X\left(G_{3,\phi\phi}+\ddot{\phi}\, G_{3,\phi X}\right)+6(2H^{2}+\dot{H})G_{4,\phi}+6H(\dot{X}+2HX)G_{4,\phi X}\nonumber \\
 &  & -6H^{2}XG_{5,\phi\phi}+2H^{3}X\dot{\phi}\, G_{5,\phi X}\,.
\end{eqnarray}

The two perfect fluids obey the following continuity equations 
\begin{eqnarray}
\dot{\rho}_{A}+3H(\rho_{A}+p_{A}) & = & 0\,,\label{rhoreq}\\
\dot{\rho}_{B}+3H(\rho_{B}+p_{B}) & = & 0\,.\label{rhomeq}
\end{eqnarray}
 Equations (\ref{be1}), (\ref{be2}), (\ref{fieldeq}), (\ref{rhoreq}),
and (\ref{rhomeq}) are not independent because of the Bianchi identities.
The field equation (\ref{fieldeq}) can be derived by using the other
equations.

\section{Conditions for the avoidance of ghosts and Laplacian instabilities
in the presence of two perfect fluids}

\label{avoidancesec}

In this section we study the stability of the flat FLRW background for the action (\ref{action}) in the presence of two perfect fluids.

Let us first consider scalar metric perturbations
$\Psi$, $\chi$, and $\Phi$ with the line element \cite{Bardeen} 
\begin{equation}
ds^{2}=-(1+2\Psi)dt^{2}+2\partial_{i}\chi\, dt\, dx^{i}+(1+2\Phi)\, d\bm{x}^{2}\,.
\end{equation}
Here we have gauged away a scalar perturbation $E$ that appears in
the form $E_{,ij}$ in front of the term $d\bm{x}^{2}$. This fixes
the spatial part of the gauge-transformation vector $\xi^{\mu}$.
In the following we choose the uniform-field gauge for which the field
$\phi$ has no perturbations, that is, $\delta\phi=0$. This fixes
the temporal part of the vector $\xi^{\mu}$.

The two perfect fluids can be described in terms of the following
Lagrangian 
\begin{equation}
S_{{\rm pf}}=\int d^{4}x\sqrt{-g}\,\left[p_{A}(\mu_{A},s_{A})+p_{B}(\mu_{B},s_{B})\right]\,,
\end{equation}
 where $\mu_{i}$ and $s_{i}$ ($i=A,B$) correspond to the chemical
potential and the entropy per particle respectively. We will employ
the method introduced in Ref.~\cite{PFs} to study the perfect fluid
from a Lagrangian point of view in order to extract the conditions
for the absence of ghost and Laplace instabilities. In the following
we will summarize and simplify the method given in Ref.~\cite{PFs}
(see also Ref.~\cite{kase}).

Since we are interested in those fluids with equations of state of
the kind $p_{i}=w_{i}\rho_{i}$ ($i=A,B$), we consider $p_{i}$ as
functions of $\mu_{i}$ alone. Here the chemical potential of each
fluid is defined as $\mu_{i}n_{i}=\rho_{i}+p_{i}$, where $n_{i}=\partial p_{i}/\partial\mu_{i}$
is the number density of the fluid $i$. In fact, it is sufficient
to give the equations of state $\mu_{i}\propto n_{i}^{w_{i}}$. We
define the fluid 4-velocity $u_{\alpha}$ associated with the chemical
potential $\mu$, as $u_{\alpha}=\mu^{-1}\partial_{\alpha}\ell$,
where $\ell$ is a scalar field. The normalization condition for the
4-velocity allows us to write the 4-velocity of the fluid $A$, as
$\mu_{A}=\sqrt{-g^{\alpha\beta}\partial_{\alpha}\ell_{A}\partial_{\beta}\ell_{A}}$.
After we perturb the field $\ell_{A}$ as $\ell_{A}(t)+\delta\ell_{A}$,
we can expand the matter Lagrangian at second order and then perform
the field redefinition $\delta\ell_{A}=-\mu_{A}v_{A}$, where $v_{A}$
is chosen to represent the independent scalar degree of freedom of
the fluid $A$. Along the same lines, the independent scalar degree
of freedom for the fluid $B$ corresponds to $v_{B}$. Since, at linear
order, the scalar fluids do not contribute to the vector perturbations,
it is sufficient to study their scalar contributions alone in order
to derive the conditions for the avoidance of ghosts and Laplacian
instabilities. We will discuss this issue in Appendix A.

We perturb the action $S=\int d^{4}x\sqrt{-g}\,{\cal L}+S_{{\rm pf}}$
up to the second order. After integrations by parts, the second-order
action is given by 
\begin{eqnarray}
S^{(2)} & = & \int d^{4}x\, a^{3}\Biggl[\bigl\{2w_{1}\dot{\Phi}-w_{2}\Psi+\sum_{l}\rho_{l}(1+w_{l})v_{l}\bigr\}\,\frac{\partial^{2}\chi}{a^{2}}+\left(\frac{1}{2}\sum_{l}\frac{(1+w_{l})\rho_{l}}{w_{l}}+\frac{w_{3}}{3}\right)\Psi^{2}+\frac{w_{4}}{a^{2}}\,(\partial\Phi)^{2}-3w_{1}\dot{\Phi}^{2}\nonumber \\
 &  & ~~~~~~~~~~~~~+\left\{ 3w_{2}\dot{\Phi}-2w_{1}\frac{\partial^{2}\Phi}{a^{2}}-\sum_{l}\frac{\rho_{l}(1+w_{l})\,(\dot{v}_{l}-3Hw_{l}v_{l})}{w_{l}}\right\} \Psi+\sum_{l}\frac{\rho_{l}(1+w_{l})}{2w_{l}}\left\{ \dot{v}_{l}^{2}-\frac{w_{l}}{a^{2}}\,(\partial v_{l})^{2}\right\} \nonumber \\
 &  & ~~~~~~~~~~~~~+3\Phi\sum_{l}\rho_{l}(1+w_{l})\,(\dot{v}_{l}-3Hw_{l}v_{l})+\frac{3}{2}\,\dot{H}\sum_{l}(1+w_{l})\rho_{l}v_{l}^{2}\Biggr]\,,\label{saction}
\end{eqnarray}
 where $l=A,B$, and 
\begin{eqnarray}
w_{1} & \equiv & 2\,(G_{{4}}-2\, XG_{{4,X}})-2X\,(G_{{5,X}}{\dot{\phi}}H-G_{{5,\phi}})\,,\label{w1def}\\
w_{2} & \equiv & -2\, G_{{3,X}}X\dot{\phi}+4\, G_{{4}}H-16\,{X}^{2}G_{{4,{\it XX}}}H+4(\dot{\phi}G_{{4,\phi X}}-4H\, G_{{4,X}})X+2\, G_{{4,\phi}}\dot{\phi}\nonumber \\
 &  & {}+8\,{X}^{2}HG_{{5,\phi X}}+2H\, X\,(6G_{{5,\phi}}-5\, G_{{5,X}}\dot{\phi}{H})-4G_{{5,{\it XX}}}{\dot{\phi}}X^{2}{H}^{2}\,,\\
w_{3} & \equiv & 3\, X(K_{,{X}}+2\, XK_{,{\it XX}})+6X(3X\dot{\phi}HG_{{3,{\it XX}}}-G_{{3,\phi X}}X-G_{{3,\phi}}+6\, H\dot{\phi}G_{{3,X}})\nonumber \\
 &  & +18\, H(4\, H{X}^{3}G_{{4,{\it XXX}}}-HG_{{4}}-5\, X\dot{\phi}G_{{4,\phi X}}-G_{{4,\phi}}\dot{\phi}+7\, HG_{{4,X}}X+16\, H{X}^{2}G_{{4,{\it XX}}}-2\,{X}^{2}\dot{\phi}G_{{4,\phi{\it XX}}})\nonumber \\
 &  & +6{H}^{2}X(2\, H\dot{\phi}G_{{5,{\it XXX}}}{X}^{2}-6\,{X}^{2}G_{{5,\phi{\it XX}}}+13XH\dot{\phi}G_{{5,{\it XX}}}-27G_{{5,\phi X}}X+15\, H\dot{\phi}G_{{5,X}}-18G_{{5,\phi}})\,,\\
w_{4} & \equiv & 2G_{4}-2XG_{5,\phi}-2XG_{5,X}\ddot{\phi}\,.\label{w4def}
\end{eqnarray}
 The equation of motion for $\chi$ gives rise to the following momentum
constraint 
\begin{equation}
w_{2}\Psi=2w_{1}\dot{\Phi}+\sum_{l=A,B}\rho_{l}(1+w_{l})v_{l}\,,\label{eq:hamconstr}
\end{equation}
 which can be used to integrate out the field $\Psi$. After replacing
$\Psi$ in Eq.~(\ref{saction}), by using Eq.~(\ref{eq:hamconstr}),
we need to integrate the terms $\partial^{2}\Phi\,\dot{\Phi}$ and
$\partial^{2}\Phi\, v_{l}$ by parts, so that the action (\ref{saction})
reduces to 
\begin{equation}
S^{(2)}=\int d^{4}x\left[{\cal A}_{ij}\dot{{\cal Q}}_{i}\dot{{\cal Q}}_{j}-{\cal C}_{ij}(\partial{\cal Q}_{i})(\partial{\cal Q}_{j})-{\cal B}_{ij}{\cal Q}_{i}\dot{{\cal Q}}_{j}-{\cal D}_{ij}{\cal Q}_{i}{\cal Q}_{j}\right],
\end{equation}
 where ${\cal Q}_{i}=(\Phi,v_{A},v_{B})$, and ${\cal A}_{ij}$, ${\cal B}_{ij}$,
${\cal C}_{ij}$, ${\cal D}_{ij}$ are the components of the $3\times3$
matrices ${\cal A}$, ${\cal B}$, ${\cal C}$, ${\cal D}$.

Imposing the matrix ${\cal A}$ to be positive definite leads to the
no-ghost conditions 
\begin{equation}
Q_{S}\equiv\frac{w_{1}(4w_{1}w_{3}+9w_{2}^{2})}{3w_{2}^{2}}>0\,,\label{Qscon}
\end{equation}
 and $w_{A}(1+w_{A})\rho_{A}>0$, $w_{B}(1+w_{B})\rho_{B}>0$. The
latter two conditions are automatically satisfied for radiation ($w_{A}=1/3$)
and non-relativistic matter ($w_{B}=0^{+}$). The speed of propagation
$c_{s}$ for the fields can be found by solving the following discriminant
equation 
\begin{equation}
\det\left(c_{s}^{2}{\cal A}-a^{2}{\cal C}\right)=0\,.\label{cs2}
\end{equation}
 This has two trivial solutions $c_{s}^{2}=w_{A}$ and $c_{s}^{2}=w_{B}$,
which are not negative for radiation and non-relativistic matter.
In order to avoid the Laplacian instability associated with the remaining
solution of Eq.~(\ref{cs2}) we require that 
\begin{equation}
c_{S}^{2}\equiv\frac{3(2w_{1}^{2}w_{2}H-w_{2}^{2}w_{4}+4w_{1}w_{2}\dot{w}_{1}-2w_{1}^{2}\dot{w}_{2})-6w_{1}^{2}[(1+w_{A})\rho_{A}+(1+w_{B})\rho_{B}]}{w_{1}(4w_{1}w_{3}+9w_{2}^{2})}\geq0\,.\label{cscon}
\end{equation}

We also consider tensor perturbations $h_{ij}$ characterized by $\delta g_{ij}=a^{2}(t)h_{ij}$,
where $h_{ij}$ is traceless ($h^{i}{}_{i}=0$) and divergence-free
($h^{ij}{}_{,j}=0$). We decompose $h_{ij}$ into the two polarization
modes, $h_{ij}=h_{\oplus}\, e_{ij}^{\oplus}+h_{\otimes}e_{ij}^{\otimes}$,
where the matrices $e_{ij}^{\oplus}$ and $e_{ij}^{\otimes}$ are
normalized to be $e_{ij}^{\lambda}(\bm{k})e_{ij}^{\lambda}(-\bm{k})^{*}=2$,
(where $\lambda={\oplus},{\otimes}$) and $e_{ij}^{\oplus}(\bm{k})e_{ij}^{\otimes}(-\bm{k})^{*}=0$
in Fourier space. The second-order action for tensor perturbations
can be written as 
\begin{equation}
S_{T}^{(2)}=\sum_{\lambda}\int d^{4}x\, a^{3}Q_{T}\left[\dot{h}_{\lambda}^{2}-\frac{c_{T}^{2}}{a^{2}}\,(\partial h_{\lambda})^{2}\right]\,,
\end{equation}
 where the conditions for the avoidance of ghosts and Laplacian instabilities
are given by 
\begin{eqnarray}
Q_{T} & \equiv & \frac{w_{1}}{4}>0\,,\label{Qtcon}\\
c_{T}^{2} & \equiv & \frac{w_{4}}{w_{1}}\geq0\,.\label{ctcon}
\end{eqnarray}

For the consistency of the theories given by the action (\ref{action})
the conditions (\ref{Qscon}), (\ref{cscon}), (\ref{Qtcon}), and
(\ref{ctcon}) need to be satisfied. Finally, as we will show in Appendix
A, the vector modes do not add any conditions to those derived above.

\section{Application to extended Galileon dark energy models}

\label{applicationsec}

In this section we shall study the cosmology based on the extended
Galileon models by taking into account the conditions for the avoidance
of ghosts and Laplacian instabilities derived in Sec.~\ref{avoidancesec}.

The covariant Galileon without the field potential \cite{DEV} corresponds
to 
\begin{equation}
K=-c_{2}X\,,\qquad G_{3}=\frac{c_{3}}{M^{3}}\, X\,,\qquad G_{4}=\frac{1}{2}\Mpl^{2}-\frac{c_{4}}{M^{6}}\, X^{2}\,,\qquad G_{5}=\frac{3c_{5}}{M^{9}}\, X^{2}\,,\label{Gali}
\end{equation}
 where $c_{i}$ ($i=2,3,4,5$) are dimensionless constants, $M_{{\rm pl}}$
is the reduced Planck mass, and $M$ is a constant having the dimension
of mass. In this case it is known that there exists a cosmological
tracker solution characterized by the condition $H\dot{\phi}=$\,constant
\cite{DePRL}. Along the tracker the dark energy equation of state
evolves as $w_{{\rm DE}}=-7/3$ (radiation era) $\to$ $w_{{\rm DE}}=-2$
(matter era) $\to$ $w_{{\rm DE}}=-1$ (de Sitter era). However, the
tracker solution for the covariant Galileon is not favored by the
joint data analysis of Supernovae Ia (SNIa), Cosmic Microwave Background
(CMB), and Baryon Acoustic Oscillations (BAO) \cite{Nesseris}. This
comes from the unusual evolution of $w_{{\rm DE}}$ during the matter
era away from $-1$. Only the solutions that approach the tracker
at late times can be allowed observationally.

Let us consider the generalization of the covariant Galileon to find
tracker solutions with different $w_{{\rm DE}}$. We take the following
functions 
\begin{equation}
K=-c_{2}M_{2}^{4(1-p_{2})}X^{p_{2}}\,,\qquad G_{3}=c_{3}M_{3}^{1-4p_{3}}X^{p_{3}}\,,\qquad G_{4}=\frac{1}{2}M_{{\rm pl}}^{2}-c_{4}M_{4}^{2-4p_{4}}X^{p_{4}}\,,\qquad G_{5}=3c_{5}M_{5}^{-(1+4p_{5})}X^{p_{5}}\,,\label{geGali}
\end{equation}
 where $c_{i}$ and $p_{i}$ ($i=2,3,4,5$) are dimensionless constants,
and $M_{i}$ ($i=2,3,4,5$) are constants having dimensions of mass%
\footnote{Kimura and Yamamoto \cite{Kimura} studied the model with $p_{2}=1,p_{3}\neq0$,
$c_{4}=0,c_{5}=0$, which recovers the Dvali-Turner model \cite{Turner}.%
}. For two perfect fluids we consider radiation (energy density $\rho_{A}=\rho_{r}$,
equation of state $w_{A}=1/3$) and non-relativistic matter (energy
density $\rho_{B}=\rho_{m}$, equation of state $w_{B}=0$). For the
choice (\ref{geGali}) the field equations (\ref{be1}) and (\ref{be2})
can be written as 
\begin{eqnarray}
 &  & 3H^{2}M_{{\rm pl}}^{2}=\rho_{{\rm DE}}+\rho_{m}+\rho_{r}\,,\label{back1}\\
 &  & (3H^{2}+2\dot{H})M_{{\rm pl}}^{2}=-p_{{\rm DE}}-\rho_{r}/3\,,\label{back2}
\end{eqnarray}
 where 
\begin{eqnarray}
\rho_{{\rm DE}} & \equiv & 2XK_{,X}-K+6H\dot{\phi}XG_{3,X}-6H^{2}\tilde{G}_{4}+24H^{2}X(G_{4,X}+XG_{4,XX})+2H^{3}\dot{\phi}X(5G_{5,X}+2XG_{5,XX})\,,\label{rhode}\\
p_{{\rm DE}} & \equiv & K-2\ddot{\phi}XG_{3,X}+2(3H^{2}+2\dot{H})\tilde{G}_{4}-12H^{2}XG_{4,X}-4H\dot{X}G_{4,X}-8\dot{H}XG_{4,X}-8HX\dot{X}G_{4,XX}\nonumber \\
 &  & -2X(2H^{3}\dot{\phi}+2H\dot{H}\dot{\phi}+3H^{2}\ddot{\phi})G_{5,X}-4H^{2}\ddot{\phi}\, X^{2}G_{5,XX}\,,\label{Pde}
\end{eqnarray}
 and $\tilde{G}_{4}\equiv G_{4}-M_{{\rm pl}}^{2}/2=-c_{4}M_{4}^{2-4p_{4}}X^{p_{4}}$.

For the covariant Galileon ($p_{2}=p_{3}=1$, $p_{4}=p_{5}=2$) there
is a tracker solution with $H\dot{\phi}=\,$constant, in which case
all the terms in Eq.~(\ref{rhode}) are proportional to $\dot{\phi}^{2}$.
We search for tracker solutions characterized by the condition 
\begin{equation}
H\dot{\phi}^{2q}={\rm constant}\,,\label{eq:deftrack}
\end{equation}
 where the power $q$ is a real constant. If we choose the following
powers, all the terms in Eq.~(\ref{rhode}) are proportional to $\dot{\phi}^{2p}$:
\begin{equation}
p_{2}=p\,,\qquad p_{3}=p+(2q-1)/2\,,\qquad p_{4}=p+2q\,,\qquad p_{5}=p+(6q-1)/2\,.\label{power}
\end{equation}
 Note that the covariant Galileon corresponds to $p=1$ and $q=1/2$.

Let us study whether the tracker really exists or not for the powers
given by Eq.~(\ref{power}). Before doing so we first discuss conditions
for the existence of a de Sitter solution characterized by $\dot{\phi}=$\,constant
and $H=$\,constant. We introduce the following variable 
\begin{equation}
x\equiv\frac{\dot{\phi}}{HM_{{\rm pl}}}\,.
\end{equation}
 For the simplification of the background equations it is convenient
to define 
\begin{eqnarray}
 &  & M_{2}\equiv(H_{{\rm dS}}M_{{\rm pl}})^{1/2}\,,\qquad M_{3}\equiv\left(\frac{{M_{{\rm pl}}}^{1-2p_{3}}}{{H_{{\rm dS}}}^{2p_{3}}}\right)^{1/(1-4p_{3})}\,,\nonumber \\
 &  & M_{4}\equiv\left(\frac{{M_{{\rm pl}}}^{2-2p_{4}}}{{H_{{\rm dS}}}^{2p_{4}}}\right)^{1/(2-4p_{4})}\,,\qquad M_{5}\equiv\left(\frac{{H_{{\rm dS}}}^{2+2p_{5}}}{{M_{{\rm pl}}}^{1-2p_{5}}}\right)^{1/(1+4p_{5})}\,,
\end{eqnarray}
 where $H_{{\rm dS}}$ is the Hubble parameter at the de Sitter solution.
For the covariant Galileon one has $M_{3}=M_{4}=M_{5}=(H_{{\rm dS}}^{2}M_{{\rm pl}})^{1/3}$.
From Eqs.~(\ref{back1}) and (\ref{back2}) we find that the de Sitter
fixed point where $\ddot{\phi}=0$ and $\dot{H}=0$ is present under
the following conditions 
\begin{eqnarray}
c_{2} & = & \frac{3}{2}\left(\frac{2}{x_{{\rm dS}}^{2}}\right)^{p}(3\alpha-4\beta+2)\,,\label{c2}\\
c_{3} & = & \frac{\sqrt{2}}{2p+q-1}\left(\frac{2}{x_{{\rm dS}}^{2}}\right)^{p+q}\left[3(p+q)(\alpha-\beta)+p\right]\,,\label{c3}
\end{eqnarray}
 where $x_{{\rm dS}}$ is the value of $x$ at the de Sitter solution,
and 
\begin{equation}
\alpha\equiv\frac{4(2p_{4}-1)}{3}\left(\frac{x_{{\rm dS}}^{2}}{2}\right)^{p_{4}}c_{4}\,,\qquad\beta\equiv2\sqrt{2}\, p_{5}\left(\frac{x_{{\rm dS}}^{2}}{2}\right)^{p_{5}+1/2}c_{5}\,.\label{ab}
\end{equation}

In order to discuss the cosmological dynamics it is convenient to
introduce the following variables 
\begin{equation}
r_{1}\equiv\left(\frac{x_{{\rm dS}}}{x}\right)^{2q}\left(\frac{H_{{\rm dS}}}{H}\right)^{1+2q}\,,\qquad r_{2}\equiv\left[\left(\frac{x}{x_{{\rm dS}}}\right)^{2}\frac{1}{r_{1}^{3}}\right]^{\frac{p+2q}{1+2q}}\,,\qquad\Omega_{r}\equiv\frac{\rho_{r}}{3H^{2}M_{{\rm pl}}^{2}}\,.\label{eq:defr1r2}
\end{equation}
 The solution (\ref{eq:deftrack}), in terms of the variable $r_{1}$,
is given by $r_{1}=1$. The de Sitter fixed point corresponds to $(r_{1},r_{2},\Omega_{r})=(1,1,0)$.
We will only consider cosmological dynamics for which $r_{1}$ and
$r_{2}$ are both positive at all times: otherwise the inverse relations
of (\ref{eq:defr1r2}) may be ill-defined. On the other hand, we will
see later on that a viable cosmology cannot allow a change of sign
for the variable $r_{2}$. Defining the dark energy density parameter
$\Omega_{{\rm DE}}\equiv\rho_{{\rm DE}}/(3H^{2}M_{{\rm pl}}^{2})$,
it follows that 
\begin{eqnarray}
\Omega_{{\rm DE}} & = & \frac{r_{1}^{\frac{p-1}{2q+1}}r_{2}}{2}\Bigl[r_{1}\!\left\{ r_{1}\bigl[12(\alpha-\beta)(p+q)+4p-r_{1}(2p-1)(3\alpha-4\beta+2)\bigr]-3\alpha(2p+4q+1)\right\} +4\beta(p+3q+1)\Bigr].\quad\;\,\label{eq:OMDE}
\end{eqnarray}
 The Friedmann equation gives the relation $\Omega_{m}\equiv\rho_{m}/(3H^{2}M_{{\rm pl}}^{2})=1-\Omega_{r}-\Omega_{{\rm DE}}$.
For the initial conditions where $r_{1}$ is positive but small ($0<r_{1}\ll1$),
we require that 
\begin{equation}
\frac{p-1}{2q+1}\geq0\,,\label{eq:pospow}
\end{equation}
 in order to have $\Omega_{{\rm DE}}\to0$. In the following we will
replace $\Omega_{m}$ with $1-\Omega_{r}-\Omega_{{\rm DE}}$.

We can obtain the differential equations for $r_{1}$, $r_{2}$, and
$\Omega_{r}$ after deriving $\ddot{\phi}$ and $\dot{H}$ from Eqs.~(\ref{be2})
and (\ref{fieldeq}), e.g., $r_{1}'/r_{1}=-2q\ddot{\phi}/(H\dot{\phi})-\dot{H}/H^{2}$,
where a prime represents a derivative with respect to $N=\ln a$.
The r.h.s.~of these differential equations can be expressed in terms
of $r_{1}$, $r_{2}$, and $\Omega_{r}$ together with the coefficients
$\alpha$, $\beta$, $p$, and $q$.

The equation for $r_{1}$ is given by 
\begin{eqnarray}
r_{1}' & = & [r_{1}(r_{1}-1)\{\beta(2p(r_{1}-1)(2r_{1}+1)-6q(r_{1}+1))+r_{1}(3\alpha(p-pr_{1}+2q)-2pr_{1})\}\nonumber \\
 &  & \times\{2((3+\Omega_{r})(2p-1)+12q)r_{1}^{(1-p)/(1+2q)}+3(4\beta(1+p+3q)-3\alpha(1+2p+4q)r_{1}\nonumber \\
 &  & +4((1+3\alpha-3\beta)p+3(\alpha-\beta)q)r_{1}^{2}+(-2-3\alpha+4\beta)(2p-1)r_{1}^{3})r_{2}\}]/\Delta\,,\label{r1eq}
\end{eqnarray}
 where 
\begin{eqnarray}
\Delta & \equiv & 2[-2r_{1}^{(2+2q-p)/(1+2q)}(3\alpha(p+2q)(2p+4q-1)-(2p+2q-1)((2+6\alpha)p+6\alpha q)r_{1}+(2+3\alpha)p(2p-1)r_{1}^{2})\nonumber \\
 &  & +4\beta r_{1}^{(1-p)/(1+2q)}(2p^{2}-p+18q^{2}+12pq-3q-3(2p+2q-1)(p+q)r_{1}^{2}+2p(2p-1)r_{1}^{3})\nonumber \\
 &  & +r_{1}^{2}\{9\alpha^{2}(2q(2qr_{1}-2q-1)+(2p^{2}+p+8pq)(r_{1}-1))(r_{1}-1)+4p^{2}r_{1}^{2}+6\alpha pr_{1}((2p+4q+1)r_{1}-2p-6q-1)\}r_{2}\nonumber \\
 &  & +3\beta^{2}(r_{1}-1)\{4p(r_{1}-1)(1+2r_{1}+6q(r_{1}+1)^{2})+4p^{2}(r_{1}-1)(1+r_{1}(2+3r_{1}))+6q(r_{1}+1)(2q(r_{1}^{2}-3)-2)\}r_{2}\nonumber \\
 &  & -3\beta r_{1}^{2}\{4pr_{1}((r_{1}-1)(2pr_{1}+1)+2q(r_{1}^{2}-2))+\alpha(r_{1}-1)(8p^{2}(r_{1}-1)(2r_{1}+1)+6q(4qr_{1}^{2}-(1+2q)r_{1}-3-10q)\nonumber \\
 &  & +4p(r_{1}-1)(12q(r_{1}+1)+r_{1}+2))\}r_{2}]\,.
\end{eqnarray}
 In addition to Eq.~(\ref{r1eq}) we write the remaining equations
of motion for the variables $r_{2}$ and $\Omega_{r}$. We find that
the second Einstein equation is equivalent to 
\begin{eqnarray}
 &  & \frac{r'_{1}}{r_{1}}\bigl\{4(3q+1)(p+2q)-2r_{2}(p+2q)r_{1}^{\frac{p-1}{2q+1}}\{3\alpha r_{1}(p-q-1)-3\beta(p-3q-2)-r_{1}^{2}[3\alpha p-3\beta(p+q)+p+3\alpha q]\}\bigr\}\nonumber \\
 &  & \qquad{}+\frac{2r'_{2}}{r_{2}}(2q+1)\bigl\{ r_{1}^{\frac{p-1}{2q+1}}r_{2}\{3\beta(p+q)+r_{1}^{2}[3\alpha p-3\beta(p+q)+p+3\alpha q]-3\alpha r_{1}(p+q)\}+2q\bigr\}\nonumber \\
 &  & \qquad{}-3(2q+1)(p+2q)\{r_{1}^{\frac{p-1}{2q+1}}r_{2}[r_{1}^{3}(4\beta-3\alpha-2)+3\alpha r_{1}-4\beta]+2\}-2(2q+1)(p+2q)\,\Omega_{r}=0\,,
\end{eqnarray}
 whereas the equation of continuity for the radiation fluid leads
to 
\begin{equation}
(2q+1)(p+2q)\,\frac{\Omega'_{r}}{\Omega_{r}}-2(3q+1)(p+2q)\,\frac{r'_{1}}{r_{1}}-2q(2q+1)\,\frac{r'_{2}}{r_{2}}+4(2q+1)(p+2q)=0\,.
\end{equation}
 These two equations, combined with Eq.~(\ref{r1eq}), completely
determine the dynamics of the system.

\subsection{Tracker}

{}From Eq.~(\ref{r1eq}) we find that there is a fixed point characterized
by 
\begin{equation}
r_{1}=1\,,
\end{equation}
 which corresponds to the tracker solution with $H\dot{\phi}^{2q}={\rm constant}$.
Considering the homogenous perturbations $\delta r_{1}$ along the
solution $r_{1}=1$, it follows that 
\begin{equation}
\delta r_{1}'=-\frac{6(p+2q)-3+(2p-1)\Omega_{r}+3r_{2}}{2(pr_{2}+2q)}\delta r_{1}\,.
\end{equation}
 For $pr_{2}+2q>0$ the tracker is stable along the $r_{1}$ direction
provided that 
\begin{equation}
6(p+2q)+(2p-1)\Omega_{r}>3(1-r_{2})\,.\label{con}
\end{equation}

Along the tracker the dark energy density parameter is simply given
by 
\begin{equation}
\Omega_{{\rm DE}}=r_{2}\,.\label{OmeDE}
\end{equation}
 For $0\le\Omega_{{\rm DE}}\le1$ the r.h.s.~of Eq.~(\ref{con})
is within the range $0\le3(1-r_{2})\le3$. As long as the condition
(\ref{con}) is satisfied, which includes the case of the covariant
Galileon, the solutions stay at the tracker. With the increase of
$r_{2}$ the solutions finally reach the de Sitter fixed point characterized
by $r_{1}=1$ and $r_{2}=1$.

Along the tracker the equations for $r_{2}$ and $\Omega_{r}$ are
given by 
\begin{eqnarray}
r_{2}' & = & \frac{(p+2q)(\Omega_{r}+3-3r_{2})}{pr_{2}+2q}\, r_{2}\,,\\
\Omega_{r}' & = & \frac{2q(\Omega_{r}-1-3r_{2})-4pr_{2}}{pr_{2}+2q}\,\Omega_{r}\,.\label{Omertra}
\end{eqnarray}
 Combining these equations, it follows that 
\begin{equation}
\frac{r_{2}'}{r_{2}}-(1+s)\,\frac{\Omega_{r}'}{\Omega_{r}}=4(1+s)\,,\qquad{\rm where}\qquad s\equiv\frac{p}{2q}\,.
\end{equation}
 Integration of this equation leads to 
\begin{equation}
r_{2}=c_{1}a^{4(1+s)}{\Omega_{r}}^{1+s}\,,\label{r2ana}
\end{equation}
 where $c_{1}$ is a constant. Since $\Omega_{r}\propto a^{-4}H^{-2}$,
the evolution of the variable $r_{2}$ is $r_{2}\propto H^{-2(1+s)}$.
Since we want $r_{2}$ to be subdominant at early times and to grow
on the tracker solution, we further impose 
\begin{equation}
1+s=1+\frac{p}{2q}>0\,.\label{eq:r2TR}
\end{equation}

Substituting Eq.~(\ref{r2ana}) into (\ref{Omertra}), we obtain
the integrated solution 
\begin{equation}
c_{1}a^{4(1+s)}{\Omega_{r}}^{1+s}=1-\Omega_{r}(1-c_{2}a)\,,\label{Omerre}
\end{equation}
 where $c_{2}$ is a constant. Therefore, the dynamics on the tracker
depends only on the free parameter $s=p/(2q)$ and the two initial
conditions $c_{1,2}$. For a particular choice of $s$ such as $s=1$
(which corresponds to the covariant Galileon), it is possible to derive
the explicit solution for $\Omega_{r}$ in terms of $a$ \cite{DeTprd,Nesseris}.
For general $s$, however, we cannot necessarily find an explicit
expression for $\Omega_{r}$. The density parameter of dark energy
along the tracker is given by $\Omega_{{\rm DE}}=r_{2}=c_{1}a^{4(1+s)}{\Omega_{r}}^{1+s}$.
Writing the density parameters of dark energy and radiation today
($a=1$) as $\Omega_{{\rm DE}}^{(0)}$ and $\Omega_{r}^{(0)}$ respectively
and using Eqs.~(\ref{OmeDE}), (\ref{r2ana}), and (\ref{Omerre}),
the coefficients $c_{1}$ and $c_{2}$ are found to be 
\begin{equation}
c_{1}=\frac{1-\Omega_{m}^{(0)}-\Omega_{r}^{(0)}}{(\Omega_{r}^{(0)})^{1+s}}\,,\qquad c_{2}=-\frac{\Omega_{m}^{(0)}}{\Omega_{r}^{(0)}}\,,
\end{equation}
 where $\Omega_{m}^{(0)}=1-\Omega_{{\rm DE}}^{(0)}-\Omega_{r}^{(0)}$
is the density parameter of non-relativistic matter today.

The dark energy equation of state $w_{{\rm DE}}\equiv p_{{\rm DE}}/\rho_{{\rm DE}}$
and the effective total equation of state $w_{{\rm eff}}\equiv-1-2\dot{H}/(3H^{2})$
along the tracker are 
\begin{equation}
w_{{\rm DE}}=-\frac{3+s(3+\Omega_{r})}{3(sr_{2}+1)}\,,\qquad w_{{\rm eff}}=-\frac{3r_{2}(s+1)-\Omega_{r}}{3(sr_{2}+1)}\,.\label{wdetra}
\end{equation}
 In the early cosmological epoch in which the condition $\Omega_{{\rm DE}}=r_{2}\ll1$
is satisfied, one has $w_{{\rm DE}}\simeq-1-s(1+\Omega_{r}/3)$ and
$w_{{\rm eff}}\simeq\Omega_{r}/3$, respectively. Hence the evolution
of $w_{{\rm DE}}$ during the radiation and matter eras is given by
$w_{{\rm DE}}=-1-4s/3$ and $w_{{\rm DE}}=-1-s$, respectively. At
the de Sitter fixed point ($r_{2}=1$ and $\Omega_{r}=0$) it follows
that $w_{{\rm DE}}=w_{{\rm eff}}=-1$. The tracker solution for the
covariant Galileon ($s=1$) is incompatible with observations because
$w_{{\rm DE}}$ is away from $-1$ during the matter and radiation
eras \cite{Nesseris}. For the compatibility with observations we
require that 
\begin{equation}
s=\frac{p}{2q}<1\,.
\end{equation}

In the regime $r_{2}\ll1$ the conditions for the avoidance of ghosts
and Laplacian instabilities are given by 
\begin{eqnarray}
\frac{Q_{S}}{M_{{\rm pl}}^{2}} & \simeq & 6q\left[p-3(\alpha-2\beta)q\right]r_{2}>0\,,\label{QS1}\\
c_{S}^{2} & \simeq & \bigl\{4p^{3}(\Omega_{r}+3)-2p^{2}\{(\Omega_{r}+3)(6\beta-3\alpha+2)-2q[3\Omega_{r}+11-3(\alpha-2\beta)(\Omega_{r}+3)]\}\nonumber \\
 &  & -3\{\beta(\Omega_{r}+3)+8q^{3}(\Omega_{r}+5)(\alpha-2\beta)-2q^{2}(7\Omega_{r}+27)(\alpha-2\beta)+q[3\alpha(\Omega_{r}+3)-2\beta(5\Omega_{r}+17)]\}\nonumber \\
 &  & -p\{(\Omega_{r}+3)(3\alpha-12\beta-1)+4q^{2}[(\alpha-2\beta)(9\Omega_{r}+33)-2(\Omega_{r}+5)]\nonumber \\
 &  & +q[12(2\beta-\alpha)(3\Omega_{r}+10)+6\Omega_{r}+22]\}\bigl\}/[24q^{2}(2p+4q-1)\{p-3(\alpha-2\beta)q\}]\geq0\,,\label{cS1}\\
\frac{Q_{T}}{M_{{\rm pl}}^{2}} & \simeq & \frac{1}{8}\left[2+3(\alpha-2\beta)r_{2}\right]>0\,,\label{QT1}\\
c_{T}^{2} & \simeq & 1-\frac{3\{2[2(\alpha-2\beta)q+3\beta]p+8(\alpha-2\beta)q^{2}+\beta(16q-3)+\beta(2p+4q-1)\Omega_{r}\}}{4q(2p+4q-1)}r_{2}\geq0\,.\label{cT1}
\end{eqnarray}
 These results reproduce those derived in Refs.~\cite{DePRL,DeTprd}
for the covariant Galileon ($p=1,q=1/2$). Let us consider the case
in which the parameters $\alpha$, $\beta$, $p$, $q$ are not very
different from the order of unity. Since $r_{2}\ll1$ the conditions
(\ref{QT1}) and (\ref{cT1}) are automatically satisfied. We see
here that $r_{2}$ cannot change its sign, as this implies the violation
of the condition $Q_{S}>0$. Since we only consider the case $r_{2}>0$
at the initial stage, the scalar ghost can be avoided for 
\begin{equation}
q\left[p-3(\alpha-2\beta)q\right]>0\,.\label{con1}
\end{equation}
 For $p$ and $q$ satisfying $q(2p+4q-1)>0$, the Laplacian instability
of the scalar perturbation is absent as long as the numerator in Eq.~(\ref{cS1})
is positive. For the covariant Galileon this corresponds to the condition
$8+10\alpha-9\beta+\Omega_{r}(2+3\alpha-3\beta)\geq0$ \cite{DePRL,DeTprd}.

\subsection{de Sitter solutions ($r_{1}=r_{2}=1$)}

We study the stability of the late-time de Sitter solution characterized
by $r_{1}=r_{2}=1$. At the de Sitter fixed point the system is effectively
described by one scalar degree of freedom $\Phi$, with the second-order
action $S^{(2)}=\int d^{4}x\, a^{3}Q_{S}[\dot{\Phi}^{2}-(c_{S}^{2}/a^{2})(\nabla\Phi)^{2}]$.
For homogenous perturbations (comoving wavenumber $k=0$) the scalar
perturbation obeys the equation of motion $\frac{d}{dt}(a^{3}Q_{S}\dot{\Phi})=0$,
whose solution is 
\begin{equation}
\Phi(t)=c_{1}+c_{2}\int^{t}\frac{1}{a^{3}Q_{S}}\, d\tilde{t}\,,\label{Phit}
\end{equation}
 where $c_{1}$ and $c_{2}$ are constants. Now we are considering
the de Sitter solution where $\dot{\phi}$ is constant, in which case
$Q_{S}$ does not vary in time. Since $a\propto e^{H_{{\rm dS}}t}$,
the second term in Eq.~(\ref{Phit}) decays in proportion to $e^{-3H_{{\rm dS}}\, t}$.
Hence the de Sitter solution in our theory is classically stable against
homogeneous perturbations.

The conditions for the avoidance of ghosts and Laplacian instabilities
(against inhomogeneous perturbations) at $r_{1}=r_{2}=1$ are given
by 
\begin{eqnarray}
\frac{Q_{S}}{M_{{\rm pl}}^{2}} & = & \frac{6(p+2q)(3\alpha-6\beta+2)[p-3(\alpha-2\beta)q]}{[2p-6(\alpha-2\beta)q-3\alpha+6\beta-2]^{2}}>0\,,\label{QS2}\\
c_{S}^{2} & = & \{6\beta+4p^{2}+p\,[9(\alpha-2\beta)^{2}+3\alpha-12\beta+4q(6\beta-3\alpha+2)-2]+3(\alpha-2\beta)[3\beta+q(9\alpha-12\beta-8q+6)]\}\nonumber \\
 &  & {}\times\frac{3(2\beta-\alpha)(2q+1)+2p-2}{6(6\beta-3\alpha-2)(p+2q)(2p+4q-1)(p-3\alpha q+6\beta q)}\geq0\,,\label{cS2}\\
\frac{Q_{T}}{M_{{\rm pl}}^{2}} & = & \frac{1}{8}\left(3\alpha-6\beta+2\right)>0\,,\label{QT2}\\
c_{T}^{2} & = & \frac{2(2p+4q-1)-3\alpha}{(2p+4q-1)(3\alpha-6\beta+2)}\geq0\,.\label{cT2}
\end{eqnarray}

\subsection{The solutions in the regime $r_{1}\ll1$ and $r_{2}\ll1$}

Equation (\ref{r1eq}) shows that there is another fixed point characterized
by 
\begin{equation}
r_{1}=0\,.
\end{equation}
 Let us study the behavior of the solutions in the regime $r_{1}\ll1$
and $r_{2}\ll1$. In doing so we consider the parameter space with
$p\ge1$ and $q>0$. Then the equations for $r_{1}$, $r_{2}$, and
$\Omega_{r}$ are approximately given by 
\begin{eqnarray}
r_{1}' & \simeq & \frac{(3+\Omega_{r})(2p-1)+12q}{2(2p+6q-1)}\, r_{1}\,,\label{r1ap}\\
r_{2}' & \simeq & \frac{(p+2q)[9-6p+\Omega_{r}(7-2p+12q)]}{2(2q+1)(2p+6q-1)}\, r_{2}\,,\label{r2ap}\\
\Omega_{r}' & \simeq & \Omega_{r}(\Omega_{r}-1)\,.\label{Omer}
\end{eqnarray}
 From Eq.~(\ref{Omer}) there are two fixed points: $\Omega_{r}=1$
and $\Omega_{r}=0$. During the radiation era ($\Omega_{r}=1$) integration
of Eqs.~(\ref{r1ap}) and (\ref{r2ap}) gives 
\begin{equation}
r_{1}\simeq a^{\frac{4p+6q-2}{2p+6q-1}}\,,\qquad r_{2}\simeq a^{\frac{(p+2q)(8-4p+6q)}{(2q+1)(2p+6q-1)}}\,,
\end{equation}
 whereas during the matter era ($\Omega_{r}=0$) one has 
\begin{equation}
r_{1}\simeq a^{\frac{3(2p+4q-1)}{2(2p+6q-1)}}\,,\qquad r_{2}\simeq a^{\frac{3(p+2q)(3-2p)}{2(2q+1)(2p+6q-1)}}\,.
\end{equation}
 The variable $r_{1}$ increases for $p\ge1$ and $q>0$, which is
followed by the approach to the tracker solution. Whether the variable
$r_{2}$ grows or not depends on the values of $p$ and $q$. If $p>3/2$
and $q>0$, for example, $r_{2}$ decreases in the matter era. However
the dynamics will in general change as the solutions approach the
tracker, $r_{1}\to1$.

The dark energy equation of state $w_{{\rm DE}}$ and the total effective
equation of state $w_{{\rm eff}}$ in the regime $r_{1}\ll1$ and
$r_{2}\ll1$ are approximately given by 
\begin{equation}
w_{{\rm DE}}\simeq\frac{1+\Omega_{r}}{2(1-2p-6q)}\,,\qquad w_{{\rm eff}}\simeq\frac{1}{3}\Omega_{r}\,.\label{wdeap}
\end{equation}
 If $p=1$ and $q=1/2$, we have $w_{{\rm DE}}=-(1+\Omega_{r})/8$
\cite{DePRL,DeTprd}.

For $(p-1)/(2q+1)>0$ the ghosts and Laplacian instabilities can be
avoided for 
\begin{eqnarray}
\frac{Q_{S}}{M_{{\rm pl}}^{2}} & \simeq & 3(p+3q)(2p+6q-1)\beta r_{1}^{(p-1)/(2q+1)}r_{2}>0\,,\label{QS3}\\
c_{S}^{2} & \simeq & \frac{p+3q-2}{2(p+3q)(2p+6q-1)}\left(1+\Omega_{r}\right)\geq0\,,\label{cS3}\\
\frac{Q_{T}}{M_{{\rm pl}}^{2}} & \simeq & \frac{1}{4}\left[1-3\beta r_{2}r_{1}^{(p-1)/(2q+1)}\right]>0\,,\label{QT3}\\
c_{T}^{2} & \simeq & 1+\frac{3(4p+12q-5-3\Omega_{r})}{4p+12q-2}\beta r_{1}^{(p-1)/(2q+1)}r_{2}\geq0\,.\label{cT3}
\end{eqnarray}
 In the regime $r_{1}\ll1$ and $r_{2}\ll1$ the conditions (\ref{QT3})
and (\ref{cT3}) are automatically satisfied.

\subsection{Other fixed points}

{}From Eq.~(\ref{r1eq}) we see that in general there are other
two more complicated fixed points, $r_{1}=r_{a,b}$, those which satisfy
the equation 
\begin{equation}
p(3\alpha-4\beta+2)r_{j}^{2}+[2\beta(p+3q)-3\alpha(p+2q)]r_{j}+2\beta(p+3q)=0\,,\label{eq:spur}
\end{equation}
 with $j=a,b$. Whether or not these fixed points are viable or not
depends on their stability and the chosen parameters of the model.

Then we will consider only those models for which either there are
no real solutions to Eq.~(\ref{eq:spur}), or, if they exist, they
are placed outside the range of interest corresponding to the interval
$0<r_{1}\leq1$. Therefore we need to fulfill the following condition
\begin{equation}
(\bar{\Delta}<0)\lor[\bar{\Delta}\geq0\land(r_{a}<0\lor r_{a}>1)\land(r_{b}<0\lor r_{b}>1)]\,,\label{eq:otherFP}
\end{equation}
 where 
\begin{eqnarray}
r_{a,b} & = & \frac{3\alpha(p+2q)-2\beta(p+3q)\pm\sqrt{\bar{\Delta}}}{2p\,(3\alpha-4\beta+2)}\,,\\
\bar{\Delta} & = & [2\beta(p+3q)-3\alpha(p+2q)]^{2}-8\beta p(3\alpha-4\beta+2)(p+3q)\,.
\end{eqnarray}

\subsection{Theoretically viable parameter space}

Let us discuss the relevant constraints we have found so far in order
to restrict the allowed parameter space. Therefore we have 
\begin{align}
\frac{p-1}{1+2q} & \geq0\,, &  & \textrm{in order to have \ensuremath{\Omega_{{\rm DE}}}subdominant et early times},\label{eq:condA}\\
1+\frac{p}{2q} & >0\,, &  & \textrm{in order to have a growing \ensuremath{\Omega_{{\rm DE}}}along the tracker },\\
\frac{p}{2q} & <1\,, &  & \textrm{in order to have the tracker consistent with data},\\
\frac{p+3q-2}{(p+3q)(2p+6q-1)} & \geq0\,, &  & \textrm{in order to have }\left.c_{S}^{2}\right|_{r_{1}\ll1,r_{2}\ll1}\geq0\,,\label{eq:condD}\\
(p+3q)(2p+6q-1)\beta & >0\,, &  & \textrm{in order to have }\left.Q_{S}\right|_{r_{1}\ll1,r_{2}\ll1}>0\,,\label{eq:condE}\\
3\alpha-6\beta+2 & >0\,, &  & \textrm{in order to have }\left.Q_{T}\right|_{{\rm de\ Sitter}}>0\,,\\
(p+2q)[p-3(\alpha-2\beta)q] & >0\,, &  & \textrm{in order to have }\left.Q_{S}\right|_{{\rm de\ Sitter}}>0\,,\\
q\left[p-3(\alpha-2\beta)q\right] & >0\,, &  & \textrm{in order to have }\left.Q_{S}\right|_{r_{1}=1,r_{2}\ll1}>0\,,\label{ds2}\\
\left.c_{S}^{2}\right|_{{\rm de\ Sitter}} & \geq0\,, &  & \!\!\left.c_{T}^{2}\right|_{{\rm de\ Sitter}}\geq0\,,\\
\left.c_{S}^{2}\right|_{r_{1}=1,r_{2}\ll1,\Omega_{r}=0} & \geq0\,, &  & \!\!\left.c_{S}^{2}\right|_{r_{1}=1,r_{2}\ll1,\Omega_{r}=1}\geq0\,.\label{eq:condM}
\end{align}

The conditions (\ref{eq:condA})-(\ref{eq:condD}) impose 
\begin{equation}
q>\frac{1}{2}\,,\quad{\rm and}\quad1\leq p<2q\,.\label{pqcon}
\end{equation}
 This implies that negative values of $p$ and $q$ are excluded.
Furthermore, the condition (\ref{eq:condE}) implies 
\begin{equation}
\beta>0\,.\label{betacon}
\end{equation}
 Altogether we find the following allowed parameter space 
\begin{eqnarray}
 &  & \left(q>\frac{1}{2}\right)\land(1\leq p<2q)\land\nonumber \\
 &  & \quad\left[\left(\frac{2p-2}{6q+3}<\alpha\leq\frac{p}{3q}\land0<\beta\leq\beta_{{\rm max}}\right)\lor\left(\frac{p}{3q}<\alpha\leq\frac{1}{3}(4p+8q-2)\land\frac{3\alpha q-p}{6q}<\beta\leq\beta_{{\rm max}}\right)\right],\label{eq:conds}
\end{eqnarray}
 where 
\begin{equation}
\beta_{{\rm max}}\equiv\frac{3\alpha-2p+6\alpha q+2}{6(2q+1)}\,.
\end{equation}
 Although the above conditions have been derived for different initial
conditions, we impose them to be true at the same time. This is because
we do not know a priori initial conditions in the early Universe,
as we have not specified how inflation works in these models.

To be more concrete, let us consider the theory with $p=1$ and $q=5/2$.
In this case the dark energy equation of state along the tracker is
given by $w_{{\rm DE}}=-1.2$ during the matter era. The condition
(\ref{eq:conds}) reduces to 
\begin{equation}
(\alpha\geq2\beta)\land\left[(\alpha>0\land\beta>0\land15\alpha\leq2)\lor\left(15\alpha>2\land\alpha<2\beta+\frac{2}{15}\land3\alpha\leq22\right)\right]\,.\label{eq:allcondspec}
\end{equation}
 Adding the condition (\ref{eq:otherFP}), the parameter space reduces
to 
\begin{eqnarray}
(\alpha\geq2\beta) & \land & \left[\left(\alpha>0\land2\sqrt{17}\sqrt{3(272-75\alpha)\alpha+68}+561\beta>408\alpha+68\land75\alpha\leq34\right)\right.\nonumber \\
 &  & \lor\left.\left(75\alpha>34\land\alpha<2\beta+\frac{2}{15}\land3\alpha\leq22\right)\right].\label{eq:allANDoth}
\end{eqnarray}

During the transition from the regime $r_{1}=1$ and $r_{2}\ll1$
to the de Sitter attractor ($r_{1}=1$ and $r_{2}=1$) the tensor
propagation speed squared $c_{T}^{2}$ can be negative, depending
on the model parameters. For $p=1$ and $q=5/2$ we have 
\begin{equation}
c_{T}^{2}=\frac{110+(22-15\alpha-99\beta-33\beta\Omega_{r})r_{2}+3(33\beta-\alpha)r_{2}^{2}}{11[2+3(\alpha-2\beta)r_{2}](5+r_{2})}\,,\label{cTex}
\end{equation}
 along the tracker. Since the transition occurs at late times, the
contribution of the radiation density parameter can be neglected in
Eq.~(\ref{cTex}). If $|\alpha|\ll1$ and $|\beta|\ll1$, then $c_{T}^{2}$
is close to 1. For positive $\alpha$ and $\beta$ of the order of
unity it happens that $c_{T}^{2}$ becomes negative. Taking the derivative
of Eq.~(\ref{cTex}) with respect to $r_{2}$, $c_{T}^{2}$ has a
minimum (the second derivative $d^{2}(c_{T}^{2})/dr_{2}^{2}$, at
$r_{2}=r_{2,{\rm min}}$, is always positive in the region $0<r_{2,{\rm min}}<1$
and when the conditions (\ref{eq:allcondspec}) are satisfied) at
\begin{equation}
r_{2,{\rm min}}=\frac{60\alpha-275\beta+3\sqrt{\Delta_{{\rm min}}}}{55\beta-12\alpha-594\beta^{2}+297\alpha\beta}\,,\label{r2so}
\end{equation}
 where 
\begin{equation}
\Delta_{{\rm min}}\equiv\beta\,[180\alpha^{2}-3\alpha(197\beta+8)+22\beta(21\beta+5)]\,.
\end{equation}
 In Eq.~(\ref{r2so}) we have chosen the solution with $r_{2}>0$,
because another solution is always negative if the conditions (\ref{eq:allcondspec})
are satisfied.

\begin{figure}
\centering\includegraphics[width=12cm]{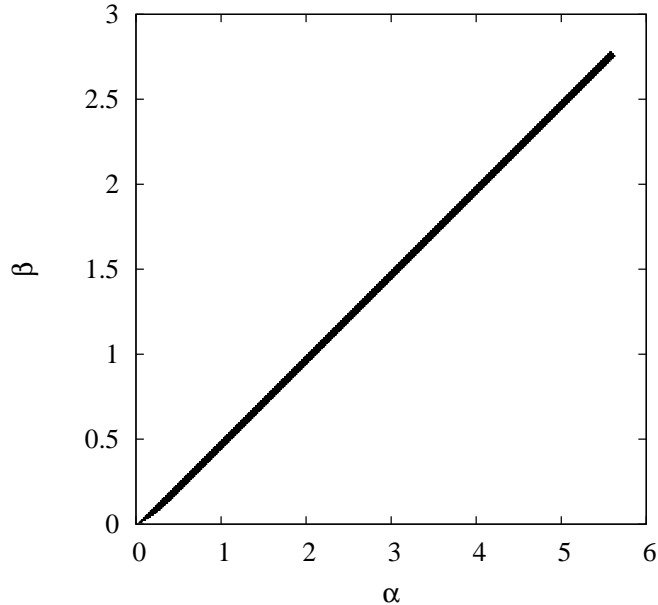} \caption{Allowed parameter space in the ($\alpha,\beta$) plane (the area in
black) for $p=1$ and $q=5/2$. We have used all the available conditions
considered so far, i.e. (\ref{eq:allANDoth}) and (\ref{eq:r2mincond}).}

\label{fig:allow} 
\end{figure}

Plugging the solution (\ref{r2so}) into Eq.~(\ref{cTex}), we find
that 
\begin{equation}
\left.c_{T}^{2}\right|_{r_{2,{\rm min}}}=\frac{36\sqrt{55}\sqrt{\Delta_{{\rm min}}}-225\alpha^{2}-15\alpha(69\beta+20)+66\beta(45\beta-23)+44}{11(30\beta-15\alpha+2)^{2}}\,.
\end{equation}
 Therefore the allowed parameter space is characterized by 
\begin{equation}
\Delta_{{\rm min}}<0\lor[\Delta_{{\rm min}}\geq0\land(r_{2,{\rm min}}\leq0\lor r_{2,{\rm min}}\geq1)]\lor[\Delta_{{\rm min}}\geq0\land(0<r_{2,{\rm min}}<1)\land\left.c_{T}^{2}\right|_{r_{2,{\rm min}}}\geq0]\,.\label{eq:r2mincond}
\end{equation}
 The first inequality comes from setting $r_{2,{\rm min}}$ to be
an imaginary number such that there is no minimum for $c_{T}^{2}$.
The second inequality allows the minimum of $c_{T}^{2}$ to be outside
the interested range for $r_{2}$.

In Fig.~\ref{fig:allow} we plot the allowed parameter space for
$p=1$ and $q=5/2$ in the $(\alpha,\beta)$ plane constrained by
the conditions (\ref{eq:allANDoth}) and (\ref{eq:r2mincond}).

\subsection{Numerical simulations}

\begin{figure}
\begin{centering}
\includegraphics[width=3.2in,height=3.3in]{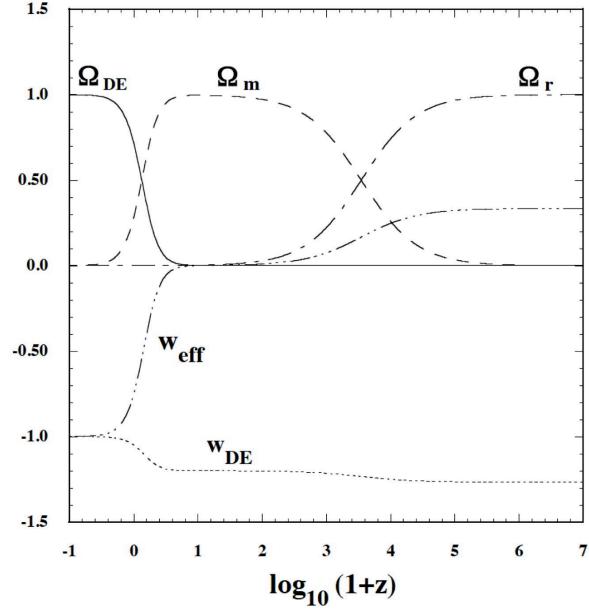} 
\par\end{centering}

\caption{Evolution of the density parameters $\Omega_{{\rm DE}}$, $\Omega_{m}$,
$\Omega_{r}$, $w_{{\rm eff}}$, and $w_{{\rm DE}}$ versus the redshift
$z$ for $p=1$, $q=5/2$, $\alpha=1$, $\beta=0.45$, and $x_{{\rm dS}}=1$.
The initial conditions are chosen to be $r_{1}=1$, $r_{2}=10^{-30}$,
and $\Omega_{r}=0.9998$ at $z=1.76\times10^{7}$. In this case the
solution is on the tracker from the beginning.}

\centering{} \label{omega} 
\end{figure}


\begin{figure}
\begin{centering}
\includegraphics[width=3.2in,height=3.2in]{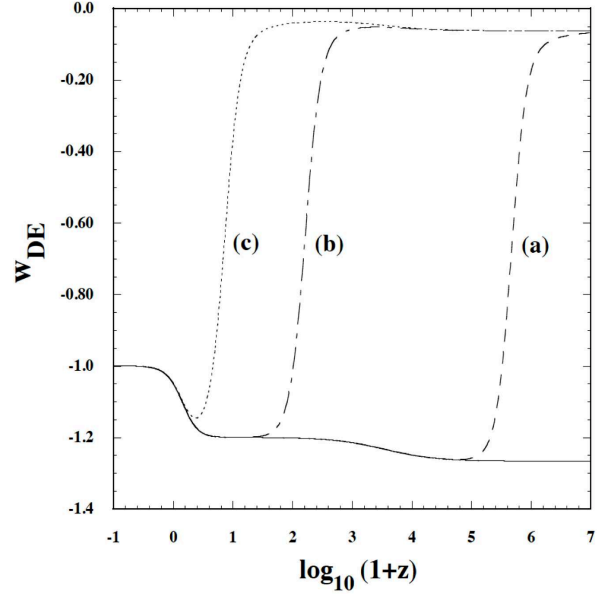} 
\par\end{centering}

\caption{Variation of $w_{{\rm DE}}$ versus $z$ for $p=1$, $q=5/2$, $\alpha=1$,
$\beta=0.45$, and $x_{{\rm dS}}=1$ with several different initial
conditions. The solid line corresponds to the tracker with the initial
conditions same as those given in Fig.~\ref{omega}. The initial
conditions for the cases (a)-(c) are (a) $r_{1}=4.0\times10^{-2}$,
$r_{2}=5.0\times10^{-26}$, $\Omega_{r}=0.9998$ at $z=1.82\times10^{7}$,
(b) $r_{1}=1.0\times10^{-5}$, $r_{2}=1.0\times10^{-13}$, $\Omega_{r}=0.9998$
at $z=1.76\times10^{7}$, and (c) $r_{1}=1.0\times10^{-7}$, $r_{2}=1.0\times10^{-9}$,
$\Omega_{r}=0.99995$ at $z=6.64\times10^{7}$, respectively.}

\centering{} \label{wde} 
\end{figure}


\begin{figure}
\begin{centering}
\includegraphics[width=3.2in,height=3in]{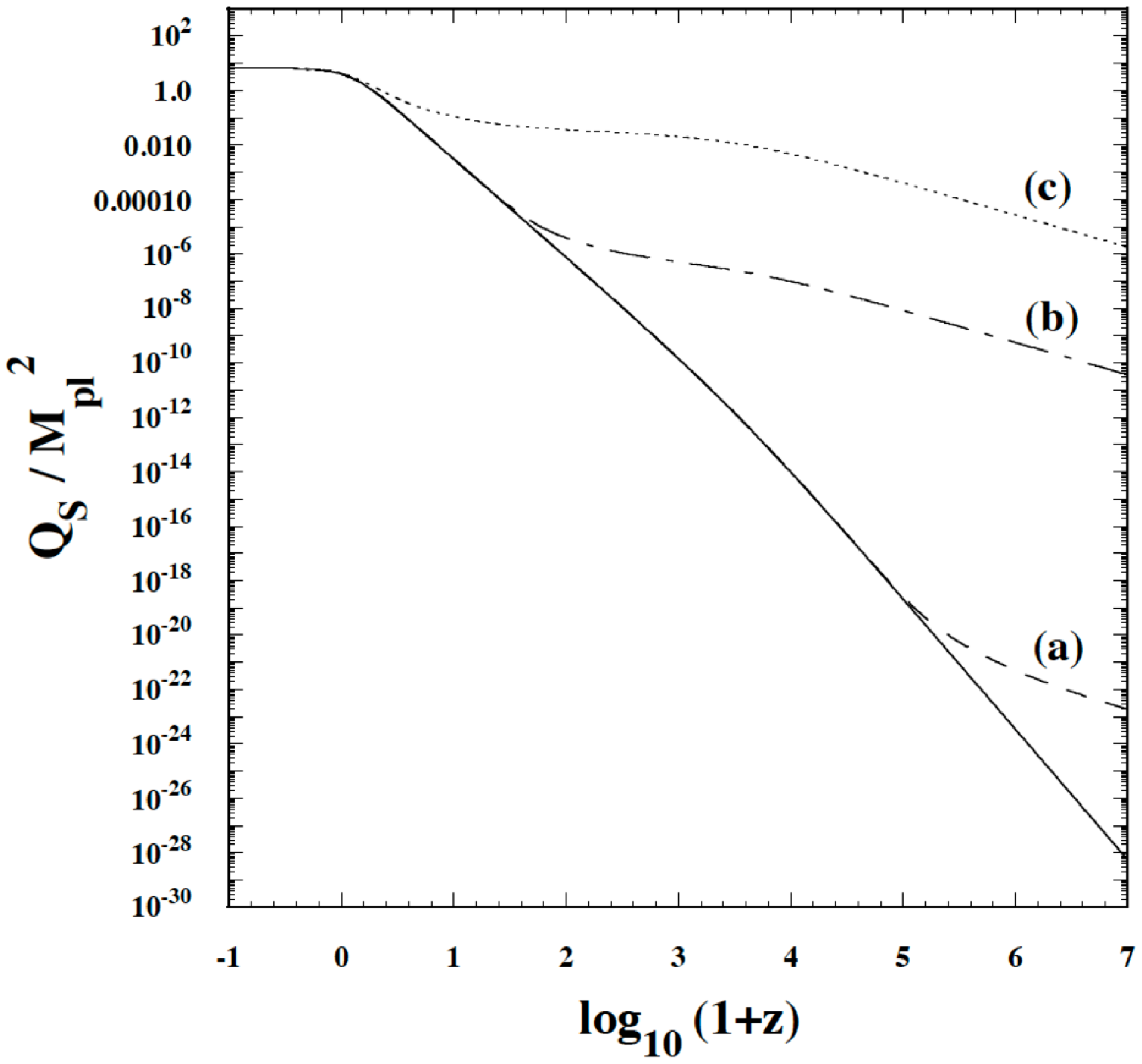} 
\par\end{centering}

\caption{Variation of $Q_{S}/M_{{\rm pl}}^{2}$ versus $z$ for the same model
parameters and initial conditions as those given in Fig.~\ref{wde}.
The solid line represents the tracker solution, whereas the cases
(a), (b), and (c) correspond to the evolution for the initial conditions
as those given in Fig.~\ref{wde}.}

\centering{} \label{sghost} 
\end{figure}


\begin{figure}
\begin{centering}
\includegraphics[width=3.2in,height=3.2in]{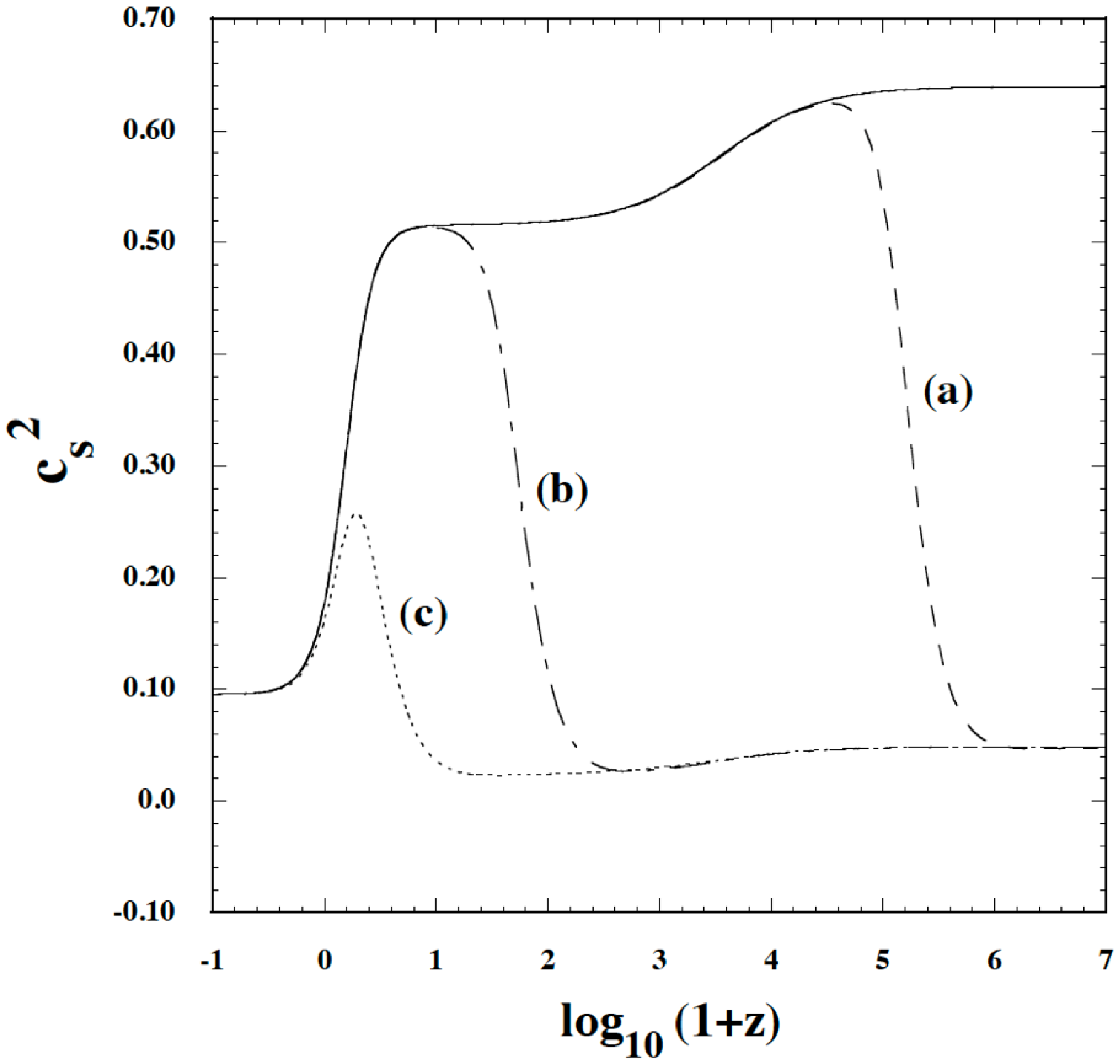} 
\par\end{centering}

\caption{Evolution of $c_{S}^{2}$ versus $z$ for the same model parameters
and initial conditions as those given in Fig.~\ref{wde}. The solid
line represents the tracker solution, whereas the cases (a), (b),
and (c) correspond to the evolution for the initial conditions as
those given in Fig.~\ref{wde}.}

\centering{} \label{cs} 
\end{figure}


In order to confirm the analytic estimation given above, we solve
the differential equations of $r_{1}$, $r_{2}$, and $\Omega_{r}$
numerically for given values of $\alpha$, $\beta$, $p$, and $q$.
In particular we study the cosmological evolution for $p=1$ and $q=5/2$,
in which case the parameters $\alpha$ and $\beta$ need to satisfy
the constraints (\ref{eq:allANDoth}) and (\ref{eq:r2mincond}).

In Fig.~\ref{omega} we plot the evolution of $\Omega_{{\rm DE}}$,
$\Omega_{m}$, $\Omega_{r}$, $w_{{\rm eff}}$, and $w_{{\rm DE}}$
versus the redshift $z=1/a-1$ for $p=1$, $q=5/2$, $\alpha=1$,
$\beta=0.45$, and $x_{{\rm dS}}=1$ with the initial conditions corresponding
to the tracker from the beginning ($r_{1}=1$). For this choice of
$\alpha$ and $\beta$ the conditions (\ref{eq:allANDoth}) and (\ref{eq:r2mincond})
are satisfied. Note that the coefficients $c_{i}$ ($i=2,\cdots,5$)
are known to be $c_{2}=9.60$, $c_{3}=18.07$, $c_{4}=4.36$, and
$c_{5}=7.20$ from Eqs.~(\ref{c2})-(\ref{ab}). Furthermore both
the fixed points given in Eq.~(\ref{eq:spur}) exist in the region
$r_{a,b}>1$. Figure \ref{omega} shows that the sequence of radiation
($w_{{\rm eff}}=1/3$, $\Omega_{r}=1$), matter ($w_{{\rm eff}}=0$,
$\Omega_{m}=1$), and de Sitter ($w_{{\rm eff}}=-1$, $\Omega_{{\rm DE}}=1$)
eras is in fact realized. The dark energy density parameter grows
as $\Omega_{{\rm DE}}=r_{2}\propto t^{2(1+s)}=t^{12/5}$ toward the
de Sitter attractor ($r_{1}=r_{2}=1$). As estimated by Eq.~(\ref{wdetra}),
the evolution of $w_{{\rm DE}}$ during the radiation and matter eras
are given by $w_{{\rm DE}}=-1.267$ and $w_{{\rm DE}}=-1.2$, respectively.
After the end of the matter-dominated epoch the solution approaches
the de Sitter attractor with $w_{{\rm DE}}=-1$.

In Fig.~\ref{wde} we show the variation of $w_{{\rm DE}}$ for the
same model parameters as those used in Fig.~\ref{omega} with a number
of different initial conditions. The approach to the tracker occurs
later for smaller initial values of $r_{1}$. This can be clearly
seen in the numerical simulations of the cases (a)-(c) in Fig.~\ref{wde}.
The case (a) corresponds to the early tracking, whereas the case (c)
to the late-time tracking with smaller initial values of $r_{1}$.
In the regime characterized by $r_{1}\ll1$ and $r_{2}\ll1$, the
analytic estimation (\ref{wdeap}) gives $w_{{\rm DE}}=-0.0625$ and
$w_{{\rm DE}}=-0.03125$ during the radiation and the matter eras,
respectively. These analytic values of $w_{{\rm DE}}$ are in good
agreement with their numerical values for the late-time tracking solution
(such as the case (c) in Fig.~\ref{wde}).

In Fig.~\ref{sghost} the evolution of the quantity $Q_{S}$ is shown
for the same model parameters and the initial conditions as those
given in Fig.~\ref{wde}. For the tracker our numerical simulations
show that $Q_{S}$ grows according to Eq.~(\ref{QS1}) in the regime
$r_{2}\ll1$ (i.e. $Q_{S}\propto r_{2}\propto t^{12/5}$), which finally
approaches the value (\ref{QS2}) at the de Sitter solution. For the
initial conditions with $r_{1}\ll1$ and $r_{2}\ll1$ we find that
the early evolution of $Q_{S}$ is well described by Eq.~(\ref{QS3}),
i.e. $Q_{S}\propto r_{2}\propto t^{19/32}$ during the radiation era.
As we can see in Fig.~\ref{sghost}, the evolution of $Q_{S}$ shifts
to that of the tracker after the solutions reach the regime around
$r_{1}=1$. Provided that $r_{2}>0$ initially, $Q_{S}$ always remains
to be positive.

As we see in Eqs.~(\ref{QT1}) and (\ref{QT3}) the quantity $Q_{T}/M_{{\rm pl}}^{2}$
is close to the value $1/4$ in the regime $r_{2}\ll1$, independently
of the values of $r_{1}$. Numerically we confirmed that, for both
the initial conditions $r_{1}=1$ and $r_{1}\ll1$, $Q_{T}/M_{{\rm pl}}^{2}$
stays the value around $1/4$ until recently and then it finally approaches
the value (\ref{QT2}) at the de Sitter solution. Provided that $3\alpha-6\beta+2>0$
the no-ghost condition for the tensor perturbation is always satisfied.

In Fig.~\ref{cs} we illustrate the evolution of $c_{S}^{2}$ for
the same model parameters and the initial conditions as those given
in Fig.~\ref{wde}. The analytic estimation (\ref{cS1}) for the
tracker gives $c_{S}^{2}=0.639$ and $c_{S}^{2}=0.515$ during the
radiation and matter dominated epochs respectively, which show good
agreement with the numerical result in Fig.~\ref{cs}. Finally $c_{S}^{2}$
approaches the value $9.48\times10^{-2}$ with a smooth transition
from the matter era to the de Sitter epoch. In the regime $r_{1}\ll1$
and $r_{2}\ll1$ the analytic estimation (\ref{cS3}) gives $c_{S}^{2}\simeq(13/544)(1+\Omega_{r})$.
In fact the numerical simulations for the cases (a), (b), and (c)
reproduce this value before the solutions reach the tracker. Since
$0<c_{S}^{2}<1$ from the radiation era to the de Sitter epoch, the
Laplacian instability of the scalar perturbation is absent.

\begin{figure}
\begin{centering}
\includegraphics[width=3.2in,height=3.2in]{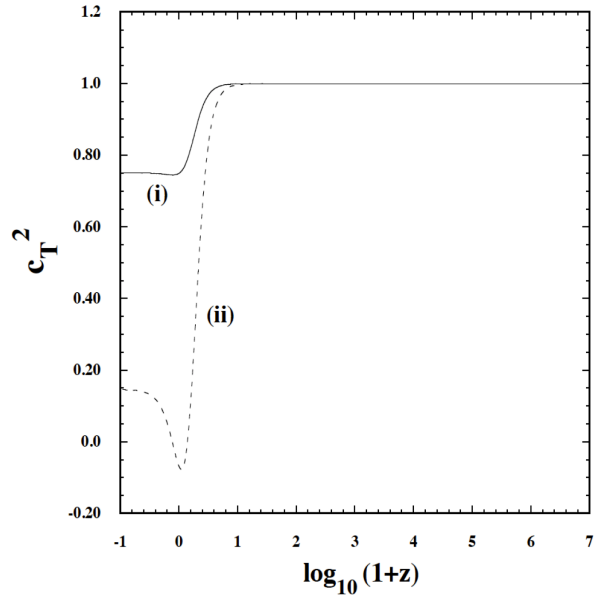} 
\par\end{centering}

\caption{Evolution of $c_{T}^{2}$ versus $z$ along the tracker for (i) $p=1$,
$q=5/2$, $\alpha=1$, $\beta=0.45$, $x_{{\rm dS}}=1$, and (ii)
$p=1$, $q=5/2$, $\alpha=6.1$, $\beta=3$, $x_{{\rm dS}}=1$. In
both cases the initial conditions are chosen to be $r_{1}=1$, $r_{2}=10^{-30}$,
and $\Omega_{r}=0.9998$. In the case (ii) the condition (\ref{eq:r2mincond})
is violated, so that $c_{T}^{2}$ temporally becomes negative during
the transition from the regime $r_{2}\ll1$ to the regime $r_{2}=1$.}

\centering{} \label{ct} 
\end{figure}


Figure \ref{ct} depicts the evolution of $c_{T}^{2}$ for the tracker
with two different combinations of $\alpha$ and $\beta$. The case
(i) corresponds to the model parameters and the initial conditions
as those given in Fig.~\ref{omega}. As estimated by Eq.~(\ref{cT1}),
the tensor propagation speed squared along the tracker is very close
to 1 but slightly less than 1 in the early cosmological epoch. In
the case (i) the model parameters satisfy the condition (\ref{eq:r2mincond}),
so that $c_{T}^{2}$ takes a positive minimum value 0.745 at $r_{2}=0.847$.
Then the solution finally approaches the de Sitter attractor with
$c_{T}^{2}=0.751$. In the case (ii), in which the condition (\ref{eq:r2mincond})
is violated, the minimum value of $c_{T}^{2}$ is negative. This shows
that the condition (\ref{eq:r2mincond}) is in fact required to avoid
the Laplacian instability of the tensor perturbation. We have also
run our numerical code for the initial conditions with $r_{1}\ll1,r_{2}\ll1$
and found that the evolution of $c_{T}^{2}$ is similar to that for
the tracker. The only difference is that the tensor propagation speed
is slightly superluminal in the regime $r_{1}\ll1,r_{2}\ll1$. In
fact, Eq.~(\ref{cT3}) shows that $c_{T}^{2}$ is slightly larger
than 1 under the conditions (\ref{pqcon}) and (\ref{betacon}).

We have also carried out the numerical simulations for other values
of $p,q$ and confirmed the accuracy of the analytic estimation. If
we choose larger values of $q$ for given $p$, the dark energy equation
of state $w_{{\rm DE}}$ along the tracker approaches $-1$. In the
limit where $q\to\infty$ the scalar propagation speed squared (\ref{cS2})
at the de Sitter solution reduces to $c_{S}^{2}\to\alpha-2\beta$
and hence we require the condition $\alpha\ge2\beta$. On the other
hand the condition (\ref{ds2}) reduces to $\alpha<2\beta$ for $q\to\infty$.
Hence, if $w_{{\rm DE}}$ along the tracker is close to $-1$, the
allowed parameter space tends to be smaller.

\section{Conclusions}

\label{consec} In the Horndeski's most general scalar-tensor theories
we derived conditions for the avoidance of ghosts and Laplacian instabilities
associated with scalar, tensor, and vector perturbations. The four
conditions (\ref{Qscon}), (\ref{cscon}), (\ref{Qtcon}), and (\ref{ctcon})
need to be satisfied for the theoretical consistency. Vector perturbations
do not give rise to any additional conditions to those derived for
scalar and tensor perturbations.

The Horndeski's action covers most of the dark energy models proposed
in literature and hence our formulas are general enough to apply them
to concrete models with second-order field equations. We proposed
new kinetically driven dark energy models described by the functions
(\ref{geGali}), which cover the covariant Galileon as a specific
case. For the choice of the powers $p_{i}$ ($i=2,3,4,5$) given in
Eq.~(\ref{power}) we showed the existence of the tracker solution
along which $H\dot{\phi}^{2q}=$\,constant. Finally the solutions
approach a stable de Sitter attractor at which $\dot{\phi}=$\,constant.

Along the tracker the dark energy equation of state during the matter
dominance is found to be $w_{{\rm DE}}=-1-p/(2q)$. The covariant
Galileon ($p=1$ and $q=1/2$) corresponds to $w_{{\rm DE}}=-2$ during
the matter era, which is not favored from the combined data analysis
of SNIa, CMB, and BAO. The extended Galileon model we proposed can
alleviate this problem because $w_{{\rm DE}}$ can be close to $-1$
for $p$ smaller than $q$.

We clarified the theoretically allowed parameter space in which the
ghosts and Laplacian instabilities are absent. For $p=1$ and $q=5/2$
we carried out numerical simulations to check the evolution of the
background quantities (like $w_{{\rm DE}}$ and $\Omega_{{\rm DE}}$)
as well as the quantities such as $c_{S}^{2}$, $c_{T}^{2}$, $Q_{S}$,
and $Q_{T}$. As we estimated analytically, the dark energy equation
of state for the tracker evolves as $w_{{\rm DE}}=-1.267$ (radiation
era), $w_{{\rm DE}}=-1.2$ (matter era), and $w_{{\rm DE}}=-1$ (de
Sitter era), see Fig.~\ref{wde}. For the initial conditions with
$r_{1}\ll1$, $w_{{\rm DE}}$ starts to evolve from the value estimated
by Eq.~(\ref{wdeap}). The approach to the tracker occurs later for
smaller initial values of $r_{1}$.

For the values of $\alpha$ and $\beta$ which are inside the allowed
parameter space, our numerical simulations show that $c_{S}^{2}$,
$c_{T}^{2}$, $Q_{S}$, and $Q_{T}$ remain to be positive in the
cosmic expansion history. Note that the condition (\ref{eq:r2mincond})
is important to avoid that $c_{T}^{2}$ becomes negative during the
transition from the matter era to the de Sitter epoch. While we showed
the cosmological evolution for one choice of $p$ and $q$, we also
confirmed that the analytic estimation is trustable for other values
of $p$ and $q$. In the limit that $p/q\to0$ the dark energy equation
of state for the tracker mimics that in the $\Lambda$CDM model.

It will be of interest to see how the combined data analysis of SNIa,
CMB, and BAO places constraints on the tracker solution in the extended
Galileon models. In order to confront this model with the observations
of large scale structure and weak lensing, we also need to study the
evolution of matter density perturbations as well as gravitational
potentials. We leave these issues for future work.

\section*{ACKNOWLEDGEMENTS}
\label{acknow} 
A.\,D.\,F.\ and S.\,T.\ were supported by the
Grant-in-Aid for Scientific Research Fund of the JSPS Nos.~10271 and 30318802. A.\,D.\,F.\ is also supported by the ThEP Center.
S.\,T.\ also thanks financial support for the Grant-in-Aid for Scientific
Research on Innovative Areas (No.~21111006). 
We thank Alexander Vikman for his initial collaboration
on this work. S.\,T.\ is grateful
to the organizers of sixth Aegean summer school and the Dark Universe
Conference for their kind hospitalities during which a part of this
work was done. 


\appendix

\appendix

\section{Ghost conditions for the vector modes\label{sec:Ghost-conditions-for}}

The no-ghost conditions for the vector modes in the presence of two
perfect fluids can be found by using the method in Ref.~\cite{PFs}.
Let us consider the perturbed metric 
\begin{equation}
ds^{2}=-dt^{2}+a\gamma_{i}dt\, dx^{i}+a^{2}(\delta_{ij}+C_{i,j}+C_{j,i})\, dx^{i}dx^{j}\,,
\end{equation}
 where $\gamma_{i,i}=0=C_{i,i}$. In order to describe the vector
perturbations at linear order, for the perfect fluid we write $\mu u_{\alpha}^{A}={\cal A}^{A}\partial_{\alpha}{\cal B}_{A}$,
where ${\cal A}^{A}$ and ${\cal B}_{A}$ are the velocity potentials
for the fluid $A$. We can choose the background values for ${\cal A}^{A}$
and ${\cal B}_{A}$ as follows: ${\cal A}^{A}=0$, and $\partial_{i}{\cal B}_{A}=b_{i}^{A}={\rm arbitrary\ constant}$.
Then, for a plane wave in Fourier space, we have $C_{i}k_{i}=0$.
By splitting ${\cal B}^{B}$ as ${\cal B}^{B}=b_{i}^{B}x^{i}+b_{i}^{B}\delta{\cal B}_{B}$,
and choosing the gauge for which $C_{\perp}=0=\delta{\cal B}_{A}$,
(where $C_{\perp}$ is the component of $C_{i}$ perpendicular to
$b_{i}^{B}$), and the arbitrary background quantities such that $b_{i}^{B}k_{i}=0$,
we find that $b_{i}^{B}$ is parallel to $C_{i}$ and both are perpendicular
to $k_{i}$.

After expanding the action at second order in the fields and integrating
out the auxiliary fields, we obtain 
\begin{equation}
S=\int d^{4}x\,\left[Q_{11}^{V}\dot{C}_{i}\dot{C}_{i}+2Q_{12}^{V}b_{i}^{B}\dot{C}_{i}\dot{\delta{\cal B}}_{B}+Q_{22}^{V}b_{i}^{B}b_{i}^{B}\dot{\delta{\cal B}}_{B}^{2}\right]\,,
\end{equation}
 where $Q_{11}$, $Q_{12}$, and $Q_{22}$ are time-dependent coefficients.
The two no-ghost conditions can be written as 
\begin{eqnarray}
Q_{11}^{V} & = & a^{5}\,{\frac{w_{{1}}{(k/a)}^{2}\,[(1+w_{{A}})\rho_{{A}}+\rho_{{B}}(1+w_{{B}})]}{2\,[2(1+w_{{A}})\rho_{{A}}+2\,\rho_{{B}}(1+w_{B})+w_{{1}}{(k/a)}^{2}]}}>0\,,\label{eq:vecghost}\\
Q_{11}^{V}Q_{22}^{V}-(Q_{12}^{V})^{2} & = & a^{10}\,{\frac{w_{{1}}{(k/a)}^{2}(1+w_{{A}})\rho_{{A}}(1+w_{{B}})\rho_{{B}}}{4[2(1+w_{A})\,\rho_{{A}}+2\,(1+w_{B})\rho_{{B}}+w_{{1}}{(k/a)}^{2}]}}>0\,.
\end{eqnarray}
 These conditions are satisfied for $w_{1}>0$ (which corresponds
to the condition (\ref{Qtcon})) and $(1+w_{{A}})\rho_{{A}}>0$, and
$(1+w_{{B}})\rho_{{B}}>0$. In General Relativity with one single
perfect fluid, only the condition (\ref{eq:vecghost}) holds, and
it agrees with the result in Ref.~\cite{PFs}, when $w_{1}=\Mpl^{2}$.
Hence the vector perturbations do not provide additional constraints
to those derived for scalar and tensor perturbations.



\end{document}